\PassOptionsToPackage{hyphens}{url}
\PassOptionsToPackage{dvipsnames,svgnames,x11names}{xcolor}
\documentclass[
  12pt]{article}

\usepackage{amsmath,amssymb,amsthm}
\usepackage{algorithm}  
\usepackage{algorithmic}
\usepackage{microtype}
\usepackage{graphicx}
\usepackage{subfigure}
\usepackage{booktabs} 
\allowdisplaybreaks[1]

\usepackage{iftex}
\ifPDFTeX
  \usepackage[T1]{fontenc}
  \usepackage[utf8]{inputenc}
  \usepackage{textcomp} 
\else 

\fi
\usepackage{lmodern}
\ifPDFTeX\else  
\fi
\IfFileExists{upquote.sty}{\usepackage{upquote}}{}
\IfFileExists{microtype.sty}{
  \usepackage[]{microtype}
  \UseMicrotypeSet[protrusion]{basicmath} 
}{}
\makeatletter
\@ifundefined{KOMAClassName}{
  \IfFileExists{parskip.sty}{%
    \usepackage{parskip}
  }{
    \setlength{\parindent}{0pt}
    \setlength{\parskip}{6pt plus 2pt minus 1pt}}
}{
  \KOMAoptions{parskip=half}}
\makeatother
\usepackage{xcolor}
\makeatletter
\ifx\paragraph\undefined\else
  \let\oldparagraph\paragraph
  \renewcommand{\paragraph}{
    \@ifstar
      \xxxParagraphStar
      \xxxParagraphNoStar
  }
  \newcommand{\xxxParagraphStar}[1]{\oldparagraph*{#1}\mbox{}}
  \newcommand{\xxxParagraphNoStar}[1]{\oldparagraph{#1}\mbox{}}
\fi
\ifx\subparagraph\undefined\else
  \let\oldsubparagraph\subparagraph
  \renewcommand{\subparagraph}{
    \@ifstar
      \xxxSubParagraphStar
      \xxxSubParagraphNoStar
  }
  \newcommand{\xxxSubParagraphStar}[1]{\oldsubparagraph*{#1}\mbox{}}
  \newcommand{\xxxSubParagraphNoStar}[1]{\oldsubparagraph{#1}\mbox{}}
\fi
\makeatother

\newtheorem{theorem}{Theorem}[section]

\newtheorem{remark}{Remark}[section]

\newtheorem{assumption}{Assumption}[section]

\usepackage{longtable,booktabs,array}
\usepackage{calc} 
\usepackage{etoolbox}
\makeatletter
\patchcmd\longtable{\par}{\if@noskipsec\mbox{}\fi\par}{}{}
\makeatother
\IfFileExists{footnotehyper.sty}{\usepackage{footnotehyper}}{\usepackage{footnote}}
\makesavenoteenv{longtable}
\usepackage{graphicx}
\makeatletter
\def\maxwidth{\ifdim\Gin@nat@width>\linewidth\linewidth\else\Gin@nat@width\fi}
\def\maxheight{\ifdim\Gin@nat@height>\textheight\textheight\else\Gin@nat@height\fi}
\makeatother
\setkeys{Gin}{width=\maxwidth,height=\maxheight,keepaspectratio}
\makeatletter
\def\fps@figure{htbp}
\makeatother

\makeatletter
\@ifpackageloaded{caption}{}{\usepackage{caption}}
\AtBeginDocument{%
\ifdefined\contentsname
  \renewcommand*\contentsname{Table of contents}
\else
  \newcommand\contentsname{Table of contents}
\fi
\ifdefined\listfigurename
  \renewcommand*\listfigurename{List of Figures}
\else
  \newcommand\listfigurename{List of Figures}
\fi
\ifdefined\listtablename
  \renewcommand*\listtablename{List of Tables}
\else
  \newcommand\listtablename{List of Tables}
\fi
\ifdefined\figurename
  \renewcommand*\figurename{Figure}
\else
  \newcommand\figurename{Figure}
\fi
\ifdefined\tablename
  \renewcommand*\tablename{Table}
\else
  \newcommand\tablename{Table}
\fi
}
\@ifpackageloaded{float}{}{\usepackage{float}}
\floatstyle{ruled}
\@ifundefined{c@chapter}{\newfloat{codelisting}{h}{lop}}{\newfloat{codelisting}{h}{lop}[chapter]}
\floatname{codelisting}{Listing}

\makeatother
\makeatletter
\@ifpackageloaded{caption}{}{\usepackage{caption}}
\@ifpackageloaded{subcaption}{}{\usepackage{subcaption}}
\makeatother

\ifLuaTeX
  \usepackage{selnolig}  
\fi
\usepackage[authoryear]{natbib}
\usepackage{bookmark}
\usepackage[marginal]{footmisc} 

\IfFileExists{xurl.sty}{\usepackage{xurl}}{} 
\hypersetup{
  pdftitle={A Statistical Theory  of Multi-stage Newton Iteration for Continual Learning},
  pdfauthor={Xinjia Lu; Chuhan Wang; Qian Zhao; Lixing Zhu; Xuehu Zhu},
  pdfkeywords={M-estimation; Continual learning；Newton iterative; Asymptotic normality},
  colorlinks=true,
  linkcolor={blue},
  filecolor={Maroon},
  citecolor={Blue},
  urlcolor={Blue},
  pdfcreator={LaTeX via pandoc}}

\newcommand{\anon}{1}

\begin{document}

\def\spacingset#1{\renewcommand{\baselinestretch}%
{#1}\small\normalsize} \spacingset{1}


\if1\anon
{
  \title{\bf Statistical Theory  of Multi-stage Newton Iteration Algorithm for Online Continual Learning}
  	\author{\small Xinjia Lu$^{1,\footnotemark[1]}$, Chuhan Wang$^{2,\footnotemark[1]}$, Qian Zhao$^{1}$, Lixing Zhu$^{2}$ and Xuehu Zhu$^{1,\footnotemark[2]}$\\
		\small$^1$ School of Mathematics and Statistics, Xi'an Jiaotong University, Xi'an, China\\
		\small $^2$ Department of Statistics, Beijing Normal University at Zhuhai, Zhuhai, China\\
	}\footnotetext[1]{* The two authors contribute equally to this work.}\footnotetext[2]{\dag Corresponding author: Xuehu Zhu. Email: zhuxuehu@xjtu.edu.cn.}
  \maketitle
} \fi

\if0\anon
{
  \bigskip
  \bigskip
  \bigskip
  \begin{center}
    {\LARGE\bf Title}
\end{center}
  \medskip
} \fi

\bigskip
\begin{abstract}
We focus on the critical challenge of handling non-stationary data streams in online continual learning environments, where constrained storage capacity prevents complete retention of
historical data, leading to catastrophic forgetting during sequential task training. 
To more effectively analyze and address the problem of catastrophic forgetting in continual learning, we propose a novel continual learning framework from a statistical perspective. Our approach incorporates random effects across all model parameters and allows the dimension of parameters to diverge to infinity, offering a general formulation for continual learning problems. To efficiently process streaming data, we 
develop a Multi-step Newton Iteration algorithm that significantly reduces computational costs in certain scenarios by alleviating the burden of matrix inversion. Theoretically, we derive the asymptotic normality of the estimator, enabling subsequent statistical inference. 
Comprehensive validation through synthetic data experiments and two real datasets analyses demonstrates the effectiveness of our proposed method.


\end{abstract}

\noindent%
\vfill

\newpage
\spacingset{1.8} 

\section{Introduction}\label{sec-intro}

With the explosive growth of data volume and the rapid development of various technologies, machine learning has undergone significant advancements and has garnered widespread attention across diverse fields and industries. Currently, continual learning \citep{chen2018lifelong, kudithipudi2022biological, wang2024comprehensive} has emerged as a pivotal and prevalent research direction in machine learning. It aims to equip machine learning algorithms with human-like continuous learning and adaptation, improving their performance and accuracy in handling increasingly complex and dynamic tasks. In the process of human learning, individuals accumulate specific knowledge and experience that can positively influence their subsequent learning endeavors \citep{french1999catastrophic, parisi2019continual, hadsell2020embracing}. Similarly, continual learning emphasizes that when faced with novel tasks, algorithms can leverage past experience to acquire new knowledge more efficiently. Concurrently, these algorithms retain a certain degree of prior knowledge, enabling them to handle previous tasks appropriately.

Traditional machine learning algorithms are generally trained and employed for fixed tasks, based on the assumption that the data conforms to an independent and identically distributed (IID) paradigm, in which samples are mutually independent and their distribution remains constant over time. These models frequently encounter difficulties in adapting to the emergence of diverse tasks and shifting data distributions. This issue is particularly prominent in the field of deep learning, where it is known as catastrophic forgetting \citep{mcclelland1995there,mccloskey1989catastrophic,french1999catastrophic,de2021continual}. More specifically, deep neural networks, which are characterized by their extensive parameter sets, necessitate considerable amounts of data for training to achieve high prediction accuracy. Although a neural network may excel in predicting outcomes for a new task following training, the modifications made to its weights and biases to optimize performance on this task often lead to a significant deterioration in its performance on previously acquired tasks.

Continual learning is one of the foremost solutions to the challenge of catastrophic forgetting. It aims to facilitate the seamless acquisition of new tasks or categories while preserving previously acquired knowledge. Within the domain of continual learning, numerous methodologies have been proposed, encompassing elastic weight consolidation \citep{kirkpatrick2017overcoming}, regularization techniques \citep{titsias2019functional}, selective experience replay \citep{isele2018selective}, dynamic network expansion \citep{yoon2017lifelong}, and Memory Aware Synapses  \citep{aljundi2018memory}. The approaches examined here primarily focus on adjusting the parameters and structure of the model, incorporating additional training data, and identifying and retaining parameter weights pertinent to previous tasks. The objective of these strategies is to mitigate the risk of catastrophic forgetting by resolving conflicts and minimizing interference between newly acquired and established knowledge.

In the current landscape of continual learning research, the majority of efforts are focused on developing innovative algorithms that can achieve superior performance on real-world datasets. Nonetheless, there is a notable scarcity of theoretical analysis, particularly within the domain of statistical analysis, concerning these algorithms and the concept of continual learning itself. \citet{zhao2024statistical} presents statistical foundations for continual learning based on regularization techniques; however, their focus is limited to linear regression tasks. Therefore, to better address the unique challenges of continual learning,  we seek to establish a more comprehensive statistical framework. This framework should naturally incorporate key characteristics of continual learning while building upon classical statistical principles.

In continual learning scenarios, different tasks typically have a common parameter structure while exhibiting task-specific variations. To effectively model these variations while preserving shared knowledge across tasks, we incorporate random effects into the model parameters. Specifically, this approach parallels random effects models in statistics, where parameters are assumed to vary randomly around a common value. However, unlike conventional random effects models where parameter variations are typically assumed to be independent of explanatory variables and attributed solely to environmental factors, our perspective differs. Building upon the setup of online continual learning tasks, we attribute these parametric variations to fundamental task disparities. Consequently, we implement a comprehensive random effects structure across all model parameters. A typical example is analogous to the one-way random effects model $\theta_k=\theta^*+\eta_k$, where $\theta_k$ varies across tasks or batches of data, and task-specific variations are captured through the random effects term $\eta_k$. 
We regard $\theta_k$ as the local optimal solution given the samples from the $k$-th task or the $k$-th batch of data, and let $\theta^*$ be the optimal solution of the global loss function, which is a weighted sum of the local loss functions. The primary objective of continual learning is to estimate this global optimum $\theta^*$ that can simultaneously address problems across all tasks. 

A motivational example involves Mendelian randomization studies.
In such studies, single nucleotide polymorphisms (SNPs) are commonly used as instrumental variables to assess the causal effect of risk factors on individual outcomes. Suppose the estimated effect of the $k$-th SNP on the risk factor is $\beta_k$, and its estimated effect on individual performance is $\gamma_k$. Then, the ratio $\gamma_k/\beta_k$ serves as a consistent estimator of the true causal effect only if the SNP qualifies as a valid instrumental variable. However, in practical scenarios, the association between the SNP and individual performance is often confounded by random confounding factors, which compromises the validity of the SNP as an instrumental variable and renders $\gamma_k/\beta_k$ an inconsistent estimator of the true causal effect. Nevertheless, we aim to identify a suitable estimator of the causal effect that can be employed to predict outcomes for new individuals. For example, \cite{bowden2016consistent} integrated data from multiple genetic variants in the Mendelian randomization framework into a single causal effect estimate using the weighted median estimator. In the numerical studies, we further analyze the MNIST dataset, where data are assigned to five tasks to simulate heterogeneous regression parameters across tasks.


Under this framework, this paper proposes a Multi-step Newton's Iterative method and provides a theoretical analysis of its statistical properties. More specifically, our main contributions can be summarized as follows:
\begin{itemize}
\item 
We developed the Multi-step Newton Iterative (MSNI) algorithm to address catastrophic forgetting in online continual learning. Compared to previous regularization-based methods, our approach is more computationally efficient for real-time data streams because it significantly reduces the number of iterations and the frequency of matrix inverse computations. Additionally, to address the inconsistency in Hessian matrix estimation caused by parameter randomness, we estimate the Hessian matrix using a larger volume of data streams than in the previous iteration. This approach ensures that the influence of the lower-precision Hessian matrix from the prior iteration on the convergence rate of the post-iteration parameter estimates becomes negligible in the asymptotic sense.

\item We established the convergence rates and asymptotic normality of the estimator, even when the parameter dimension grows at a certain rate with the sample size. These results can be the base for performing hypothesis testing and constructing confidence intervals. More specifically, we derive the following results:
(1) For heterogeneous data streams (where parameters vary across batches), 
we provide the first-ever derivation of the asymptotic normality of the MSNI statistics.
(2) For homogeneous data streams (where parameters remain constant across batches),
our algorithm achieves the optimal  convergence rate for parameter estimates. In this case, the statistical properties of our estimator are identical to those of a weighted least squares distributed computing method (\cite{zhu2021least}), as if all samples were processed together.

\end{itemize}

\section{Related Work}
Recently, there has been noteworthy investigative work conducted on continual learning \citep{de2021continual, wang2024comprehensive}, and in this section, we provide a brief overview of some of the relevant research in this field. The field of continual learning encompasses numerous scenarios, with Instance-Incremental Learning (IIL) \citep{lomonaco2017core50} representing the most straightforward among them. In IIL, there is a single task, and the data is received in batches. According to the characteristics of the data distribution and labeling space pertinent to the task to be processed, there are broadly three primary types of continual learning scenarios \citep{hsu2018re, van2019three}: Domain-Incremental Learning (DIL) involves scenarios where the distribution of data varies across different tasks, yet the label space remains consistent; Task-Incremental Learning (TIL) refers to scenarios in which the data distribution varies across tasks, while the task label spaces are disjoint, but the specific type of task to be predicted remains known throughout; Class-Incremental Learning (CIL) differs from TIL in that the target prediction task is unspecified, necessitating the identification of the correct category within a potentially vast set of categories for categorization tasks.

In recent years, numerous effective continual learning methods have been proposed to address the problem of catastrophic forgetting.
Experience replay-based continual learning methods employ a small buffer to store previously trained samples, and the selection and utilization of these stored samples become crucial factors in their effectiveness \citep{riemer2018learning, caccia2020online}. \citet{chaudhry2019tiny} analyzes the advantages of utilizing tiny historical datasets for continual learning processes. Addressing the issue of catastrophic forgetting through the enhancement of optimization procedures \citep{zeng2019continual, farajtabar2020orthogonal, wang2021training, tang2021layerwise, lin2022beyond} represents another significant research avenue. Continual learning approaches that rely on regularization techniques \citep{kirkpatrick2017overcoming, zenke2017continual, liu2018rotate} emphasize the iterative refinement of the Hessian matrix via diverse approximation methodologies. 
\citet{ritter2018online} exhibits superior performance compared to both EWC and SI by leveraging the Kronecker factored online Laplace approximation. \citet{lee2020continual} proposes a continual learning method that incorporates quadratic penalization by extending the Kronecker-factored approximate curvature. 
\citet{wu2024meta} integrates Meta-Continual learning with these methods to develop a more efficient strategy and proposes a framework for regularization-based continual learning approaches.

The aforementioned approaches typically begin with either Taylor expansion of the loss function or employing Bayesian theorem to analyze conditional probability density functions, ultimately focusing on methodological solutions to catastrophic forgetting. While these methodological contributions are valuable, they lack a rigorous theoretical analysis from a classical statistical perspective. 
Therefore, this paper establishes a statistical framework for continual learning, aiming to provide a more systematic and theoretically grounded statistical analysis of the problem, and we propose a Multi-step Newton Iteration algorithm. It is worth noting that  \citet{zhao2024statistical} conducts a statistical-theoretical analysis of regularization-based continual learning methods; however, their modeling framework was restricted to linear regression tasks while lacking investigation into asymptotic statistical properties. The present work bridges this theoretical gap by establishing relevant results under a more general framework, including the derivation of a weakly normal limiting distribution.

\section{Preliminaries}

\subsection{Notation}

Firstly, we provide some notations.  $v$ is a $p$-dimensional constant vector. For any vector $v=(v_1,\ldots,v_p)^{\top}\in \mathbb{R}^p$, define $\|v\|=(\sum_{j=1}^{p}v_j^2)^{1/2}$ and $v^{\otimes 2}=vv^{\top}$.
Let the closed ball $B(v,r)=\{w\in\mathbb{R}^p:\|w-v\|\leq r\}$. 
For any positive constant $N$, let $[N]$ be the index set $\{1,2,\ldots,N\}$, and $\lfloor N\rfloor$ be the largest interger no more than $N$. 
For a matrix $A$, denote $\|A\|=\sup\limits_{\|v\|=1}\|v^{\top}A\|$ as the spectral norm, which is equal to the largest eigenvalue of $A$. In addition, $\lambda_{\max}(A)$  and $\lambda_{\min}(A)$ represent  the largest eigenvalue and the smallest eigenvalue of $A$, respectively. 
Denote $\mathbf{N}(0,1)$ be the standard normal distribution. Use $\mathcal{A}^c$ to represent the complementary of event $\mathcal{A}$. We say $X_n=O_p(r_n)$ if for any $\epsilon>0$, there exists a positive constant $M$ such that $P(|X_n/r_n|<M)>1-\epsilon$. 
For the function $l(X,Y,\theta)$ with the parameter $\theta=(\theta_1,\ldots,\theta_p)^{\top}$,  let $\dot{l}(X,Y,\theta)=\frac{\partial l(X,Y,\theta)}{\partial\theta}$, $\ddot{l}(X,Y,\theta)=\frac{\partial^2 l(X,Y,\theta)}{\partial\theta^2}$, $\dot{l}_j(X,Y,\theta)=\frac{\partial l(X,Y,\theta)}{\partial\theta_j}$ and $\ddot{l}_{j_1j_2}(X,Y,\theta)=\frac{\partial^2 l(X,Y,\theta)}{\partial\theta_{j_1}\partial\theta_{j_2}}$. 

\subsection{Problem Setup}

In the online continual learning setting, there are a total of $M$ tasks, and the data for each task arrives in batches. We assume the data consists of $K$ batches $\{X_k,Y_k\}=\{X_{(k,i)},Y_{(k,i)}\}_{i=1}^{n_k}, k=1,2,\cdots,K$, where each batch contains $n_k$ observations of $p$-dimensional random vectors $X_{(k,i)} \in \mathbb{R}^p$ with responses $Y_{(k,i)}, i=1,2,\cdots,n_k$. The data for each task is independent and the task types are consistent across different scenarios, but their target parameters may vary to some extent. We assume that the data is generated by the following model:
\begin{align}\label{eq1}
Y_{(k,i)}=h(X_{(k,i)},\theta_k)+\varepsilon(\theta_k),
\end{align}
where $h$ is the regression function satisfying $h(X_k,\theta_k)=E(Y_k|X_k)$, $\varepsilon(\theta_k)$ is the error term with zero mean on the $k$-th data stream. For the same task, the parameters $\theta_k$ are also the same, i.e., for fixed $k$, $\theta_k$ does not change with $i$ in (\ref{eq1}). In our setup, the data for each task arrives in online batches. The data for task $1, 2, \cdots,M$ arrive sequentially in batches, and each batch of data corresponds to a specific task. Since we process data in batches, we will use batch indices $k$ instead of task indices $T$ for ease of analysis in the following discussion.

For each batch of data, we have the loss function $l(X_k,Y_k,\theta_k)$ and
\begin{align*}
\theta_k
=&\operatorname{argmin}\limits_{\theta\in\mathbf{\Theta}}\mathbb{E}_{X_k,Y_k}\{l(X_k,Y_k,\theta)\}\\
=&\operatorname{argmin}\limits_{\theta\in\mathbf{\Theta}}\int_{X_k,Y_k}l(x,y,\theta)p_k(x,y) dxdy,
\end{align*}
where $l(\cdot,\cdot,\cdot)$ represents the loss function, $p_k(x,y)$ is the density function of $\{X_k,Y_k\}$, and $\theta_k$ is the local optimal solution corresponding to the $k$-th batch of data samples. By weighting the above local loss function with respect to the parameters $\theta$, we can obtain the global loss function. 

To quantify the parameter discrepancy, we introduce $\Delta(\theta)$ instead of directly using $\theta$, with the detailed interpretation as follows. For any $\theta\in\mathbf{\Theta}$, define $\Delta(\theta)=\theta_k-\theta$. Then $\Delta(\theta)$ is a random variable and the distribution of $\{X,Y\}$ only depends on $\Delta(\theta)$ (regard $\theta$ as a constant vector). For notational convenience, given any observed data $\{X,Y\}$, we denote its density function by $p_{\Delta(\theta)}(x,y)$. Similarly, the density function of $\Delta(\theta)$ is denoted as $p_{\theta}(\Delta)$. 
Define
\begin{align*}
F(\theta)=&\int_{\Delta(\theta)}\left[\int_{X,Y}l(x,y,\theta)
    p_{\Delta(\theta)}(x,y)dxdy\right]p_{\theta}(\Delta)d\Delta
    \\=&\mathbb{E}_{\Delta(\theta)}[\mathbb{E}_{X,Y}\{l(X,Y,\theta)|\Delta(\theta)\}].
\end{align*}
Here $F(\cdot)$ is the global loss function, which consists of two expected values. The inner expectation ($\mathbb{E}_{X,Y}$) quantifies the expected loss over the samples in a single data stream, given a fixed parameter value for that stream, as the sample size tends to infinity. The outer expectation ($\mathbb{E}_{\Delta(\theta)}$) captures the average loss per data stream in expectation, as the number of data streams  diverges to infinity. Meanwhile, the empirical risk version of $F(\theta)$ can be expressed as: $$\lim\limits_{K\to\infty}\frac{1}{K}\sum_{k=1}^K\lim\limits_{n_k\to\infty}\frac{1}{n_k}\sum_{i=1}^{n_k}l(X_{k,i},Y_{k,i},\theta).$$
Our objective is to estimate the optimal parameter that minimizes the global loss function, given a sufficiently large number of data streams and adequate sample sizes within each stream, i.e., to find
\begin{align*}
\theta^*=\operatorname{argmin}\limits_{\theta\in\mathbf{\Theta}}F(\theta)=\operatorname{argmin}\limits_{\theta\in\mathbf{\Theta}}\lim\limits_{K\to\infty}\frac{1}{K}\sum_{k=1}^K\lim\limits_{n_k\to\infty}\frac{1}{n_k}\sum_{i=1}^{n_k}l(X_{k,i},Y_{k,i},\theta).
\end{align*}
Obviously, the above objective can be approximated by minimizing the following empirical risk function:
\begin{align}\label{2}
    \hat{\theta}=\operatorname{argmin}\limits_{\theta\in\mathbf{\Theta}}\frac{1}{K}\sum_{k=1}^{K}\frac{1}{n_k}\sum_{i=1}^{n_k}l(X_{(k,i)},Y_{(k,i)},\theta).
\end{align}
For ease of notations, we define
\begin{align*}
L_k(\theta)
:=\frac{1}{n_k}\sum_{i=1}^{n_k}l(X_{(k,i)},Y_{(k,i)},\theta),
\end{align*}
 then $\hat{\theta}=\operatorname{argmin}\limits_{\theta\in\mathbf{\Theta}}\frac{1}{K}\sum_{k=1}^KL_k(\theta)$. Note that for a single batch of data, the parameter can be estimated via optimizing the loss function on a single batch of data as
$\hat{\theta}_k=\operatorname{argmin}\limits_{\theta\in\mathbf{\Theta}}L_k(\theta)$.
Thus, it is natural to perform a second-order Taylor expansion of $L_k(\theta)$ at $\hat{\theta}_k$:
\begin{align*}
    	L_k(\theta)
	=&\frac{1}{n_k}\sum_{i=1}^{n_k} \left\{l(X_{(k,i)},Y_{(k,i)},\theta)-l(X_{(k,i)},Y_{(k,i)},\hat{\theta}_k)\right\} +C_1\\
	\approx& \frac{1}{n_k}\sum_{i=1}^{n_k}(\theta-\hat{\theta}_{k})^{\top}\ddot{l}(X_{(k,i)},Y_{(k,i)},\hat{\theta}_k)(\theta-\hat{\theta}_{k})+C_2\\
	=&(\theta-\hat{\theta}_{k})^{\top}\left\{\frac{1}{n_k}\sum_{i=1}^{n_k}\ddot{l}(X_{(k,i)},Y_{(k,i)},\hat{\theta}_k)\right\}(\theta-\hat{\theta}_{k})+C_2\\
	=&(\theta-\hat{\theta}_{k})^{\top}\{\hat{\Sigma}^{-1}_k\}(\theta-\hat{\theta}_{k})+C_2,
\end{align*}
where $\hat{\Sigma}^{-1}_k=\frac{1}{n_k}\sum_{i=1}^{n_k}\ddot{l}(X_{(k,i)},Y_{(k,i)},\hat{\theta}_k)$, and $C_1$ and $C_2$ are positive constants unrelated to $\theta$.
If the difference between $\hat{\theta}_k$ is neglected, then we can estimate $\hat{\theta}$ in (\ref{2}) by minimizing the following equation
$$\operatorname{argmin}\limits_{\theta\in\mathbf{\Theta}}\frac{1}{K}\sum_{i=1}^K(\theta-\hat{\theta}_{k})^{\top}\{\hat{\Sigma}^{-1}_k\}(\theta-\hat{\theta}_{k}),$$
this immediately leads to a weighted least squares estimate (WLSE)
$$\hat{\theta}_{WLSE}=\left(\sum_{k=1}^K\hat{\Sigma}^{-1}_k\right)^{-1}\left(\sum_{k=1}^{K}\hat{\Sigma}^{-1}_k\hat{\theta}_k\right). $$
This estimation method has been successfully employed in the treatment of distributed problems \citep{zhu2021least}. In continual learning scenarios, the approach demonstrates reasonable effectiveness when task similarity is high, but its performance degrades as inter-task divergence increases. Recent research in continual learning has provided detailed analyses of parameter variation effects similar to those described \citep{hochreiter1997flat, keskar2016large, mirzadeh2020understanding}. Consequently, we introduce a Multi-step Newton Iteration(MSNI) method designed to achieve enhanced performance in continual learning problems, accompanied by a rigorous analysis of its statistical properties.

\section{Multi-step Newton Iterative algorithm}

In order to solve the challenge of continual learning in the presence of large variations in task parameters, in this section we present a Multi-step Newton Iterative (MSNI) algorithm that exploits the gradient and Hessian matrix information of the data to cope with catastrophic forgetting problem. To establish the theoretical foundations of the proposed algorithm, we first present several assumptions before detailing the algorithmic framework and its theoretical analysis.

\subsection{Assumptions}
\begin{assumption}(Convexity)\label{a11}
	The loss function $l(X,Y,\theta)$ is convex with respect to $\theta\in\mathbf{\Theta}$.
\end{assumption}
\begin{assumption}\label{a1}(Bounded eigenvalue)
	There exists a positive constant $\lambda$ such that for any $\theta\in\mathbf{\Theta}$, $\lambda^{-1}\leq \lambda_{\min}(\mathbb{E}\{\ddot{l}(X,Y,\theta)\})\leq \lambda_{\max}(\mathbb{E}\{\ddot{l}(X,Y,\theta)\})\leq \lambda$.
\end{assumption}
\begin{assumption}\label{a3}(Lipschitz condition) For any $\theta,\theta{'}\in\mathbf{\Theta}$ , we have
	\begin{align*}
		\|\ddot{l}(X,Y,\theta)-\ddot{l}(X,Y,\theta{'})\|\leq C(X,Y)\|\theta-\theta{'}\|,
	\end{align*}
	where $\mathbb{E}\{C(X,Y)^2\}\leq C_1^2$.
\end{assumption}
\begin{assumption}\label{a4}(Sub-exponenital for the gradient) For any $\theta\in\mathbf{\Theta}$ and $j\in[p]$, there exists a constant $s$ such that $\mathbb{E}[\exp\{s|\dot{l}_j(X,Y,\theta)|\}]\leq 2$. 
\end{assumption}
\begin{assumption}\label{a6}(Moment restriction)
	For any $\theta\in\mathbf{\Theta}$ and  $j_1,j_2\in[p]$,  $\mathbb{E}\{\ddot{l}_{j_1,j_2}(X,Y,\theta)^2\}\leq C_3$, where $C_3$ is a positive interger.
\end{assumption}
\begin{assumption}\label{ab}(Berry-Esseen condition 1) For any constant vector $v=(v_1,v_2,\cdots,v_p)^{\top}$, there exists a positive constant $C_{BE}$  such that
    \begin{align*}
		\frac{\mathbb{E}[|v^{\top}\mathbb{E}\{\dot{l}(X_{(k,i)},Y_{(k,i)},\theta^*)|\Delta(\theta^*)=\theta_k-\theta^*\}|^3]^{2/3}}{\mathbb{E}[\{v^{\top}\mathbb{E}\{\dot{l}(X_{(k,i)},Y_{(k,i)},\theta^*)|\Delta(\theta^*)=\theta_k-\theta^*\}\}^2]}\leq C_{BE}.
	\end{align*}
\end{assumption}
\begin{assumption}\label{ac}(Berry-Esseen condition 2) For any constant vector $v=(v_1,v_2,\cdots,v_p)^{\top}$, there exists a positive constant $C'_{BE}$  such that
	\begin{align*}
		\frac{\mathbb{E}[|v^{\top}\dot{l}(X_{(k,i)},Y_{(k,i)},\theta^*)|^3]^{2/3}}{\mathbb{E}[\{v^{\top}\dot{l}(X_{(k,i)},Y_{(k,i)},\theta^*)\}^2]}\leq C'_{BE}.
	\end{align*}
\end{assumption}

The above conditions are commonly used conditions in research on M-estimation.
Assumption \ref{a11} assumes that the parameter space is convex, which has been used in numerous studies related to M-estimation such as  \citet{van2000asymptotic}. In Newton's iterative algorithm, the global convexity condition of the loss function can be replaced by the consistency assumption of the initial iteration value and local strong convexity (the second derivative of loss function is positive definite in the neighborhood of $\theta^*$). If $l_k(X,Y,\theta)$ are strongly convex near $\theta^*$, then the sum of $L_k$ is obviously also strongly convex near $\theta^*$. Assumption \ref{a1} restricts the eigenvalues of the Hessian matrix, and similar assumptions can be found in  \citet{zhang2012communication}. Assumption \ref{a3} is a Lipschitz condition, which has been used by  \citet{jordan2019communication}. Assumption \ref{a4} stipulates that each element of the gradient follows a sub-exponential distribution. Assumption \ref{a6} is a general moment restriction for the second-order derivative. Finally,  Assumptions \ref{ab} and \ref{ac} are Berry-Esseen conditions, which are used to ensure Central Limit Theorem. 

\subsection{One-stage Newton Iteration(OSNI)}

Firstly, we give a One-step Newton Iterative algorithm for a simple case. As the data stream gradually arrives, to provide a good initial estimate for the iterative algorithm, we first store some batches of data and then solve for an M-estimate, which minimizes the following objective function:
\begin{align}\label{4}
\hat{\theta}_{stage,0}=\operatorname{argmin}\limits_{\theta\in\mathbf{\Theta}}\frac{1}{\lfloor K^{\alpha}\rfloor}\sum_{k=1}^{\lfloor K^{\alpha}\rfloor}\frac{1}{n_k}\sum_{i=1}^{n_k}\dot{l}(X_{(k,i)},Y_{(k,i)},\theta),
\end{align}
and then obtain the initial estimate $\hat{\theta}_{stage,0}$, where $\alpha \in (0,1)$ and $\lfloor K^{\alpha}\rfloor$ is a positive integer that remains relatively small due to storage constraints. For existing data and upcoming data, we use this initial estimation to calculate the gradient and Hessian matrix information:
$$\dot{L}_k(\hat{\theta}_{stage,0})=\frac{1}{n_k}\sum_{i=1}^{n_k}\dot{l}(X_{(k,i)},Y_{(k,i)},\hat{\theta}_{stage,0})\quad\text{and}\quad   \ddot{L}_k(\hat{\theta}_{stage,0})=\frac{1}{n_k}\sum_{i=1}^{n_k}\ddot{l}(X_{(k,i)},Y_{(k,i)},\hat{\theta}_{stage,0}).$$
Finally we perform a Newton iteration, 
\begin{equation}\label{5}
    \begin{aligned}
            \hat{\theta}_{stage,1}=\hat{\theta}_{stage,0}
    -\left\{\frac{1}{\lfloor K^{\alpha_1}\rfloor}\sum_{k=1}^{\lfloor K^{\alpha_1}\rfloor}\ddot{L}_k(\hat{\theta}_{stage,0})\right\}^{-1}\left\{\frac{1}{\lfloor K^{\alpha_1}\rfloor}\sum_{k=1}^{\lfloor K^{\alpha_1}\rfloor}\dot{L}_k(\hat{\theta}_{stage,0})\right\},
    \end{aligned}
\end{equation}
which we show specifically in Algorithm \ref{alg2}.

\begin{algorithm}[ht]
   \caption{One-stage Newton Iteration(OSNI)}
   \label{alg2}
\begin{algorithmic}
   \STATE {\bfseries Input:} $\{X_{(k,i)},Y_{(k,i)} \}_{i=1}^{n_k},k=1,\cdots,K$, $0<\alpha<\alpha_1\leq 1$.
   \STATE {\bfseries Output:} $\hat{\theta}_{stage,1}.$
   \STATE Obtain $\hat{\theta}_{stage,0}$ by (\ref{4}).
   \FOR{$k=1$ {\bfseries to} $\lfloor K^{\alpha_1}\rfloor$}
   \STATE Calculate and store the gradient $\dot{L}_k(\hat{\theta}_{stage,0})$.
   \STATE Calculate and store the Hessian matrix  $\ddot{L}_k(\hat{\theta}_{stage,0})$.
   \ENDFOR
    \STATE Calculate $\hat{\theta}_{stage,1}$ by (\ref{5}).
\end{algorithmic}
\end{algorithm}

Specifically, we compute the initial estimate $\hat{\theta}_{stage,0}$ along with its gradient and Hessian matrix using only the first $\lfloor K^{\alpha}\rfloor$ data streams. After processing these streams, all raw data are permanently discarded. For each subsequent data stream, we update the gradient and Hessian aggregates without retaining any raw observations. Finally, upon receiving all $K$ streams, we refine the initial estimate through a single Newton-Raphson iteration using the accumulated gradient and Hessian information.

In Algorithm \ref{alg2}, we use the Hessian matrix to restrict the direction of parameter iteration updates. The gradient allows the parameters to be updated towards the target direction of the new task, while the inverse of the Hessian matrix ensures that the updated parameters do not deviate too much from the previous task. This is consistent with the regularization based continual learning methods. However, in this process, we consistently use the old parameters to compute gradients and Hessian matrices, making the difference between the initial estimate and the true parameters extremely important. Therefore, Algorithm \ref{alg2} works well when the gap between the tasks is small; however, when the gap between the tasks is large, there is a high demand on the amount of data to be stored initially. 

To further elucidate these properties, we subsequently present theoretical analyses and establish the consistency of our initial estimator.
\begin{theorem}\label{thm1}
	Under Assumptions \ref{a11}-\ref{a6}, if $p/K^{\alpha}=o(1)$, we have 
	\begin{align*}
		\|\hat{\theta}_{stage,0}-\theta^*\|=O_p(\sqrt{p/K^{\alpha}}).
	\end{align*}
\end{theorem}
\begin{remark}
Theorem \ref{thm1} establishes the convergence rate of the estimator when we only take the advantage of the information from $K^{\alpha}$ tasks. This provides a theoretical guarantee for subsequent Newton-Raphson iterations based on this initial estimator. From Theorem \ref{thm1}, we know that $\hat{\theta}_{stage,0}$ achieves a convergence rate of $\sqrt{p/K^{\alpha}}$ toward the true parameter $\theta^*$, highlighting its asymptotic behavior under the given computational constraints. 
\end{remark}

Building on the consistency guarantee of Theorem \ref{thm1}, we now analyze the estimator in Algorithm \ref{alg2}. Theorem \ref{thm2} characterizes the convergence rate of the updated estimator after one-step Newton-Raphson iteration starting from the initial parameter estimate. It further specifies conditions under which the estimator achieves asymptotic normality. 

\begin{theorem}\label{thm2}

	Let $\alpha_1=1$.  Under Assumptions \ref{a11}-\ref{a6}, 
    if $pK/K^{2\alpha}=o(1)$, then $\|\hat{\theta}_{stage,1}-\theta^*\|$ reaches the optimal rate $O_p(\sqrt{p/K})$. Further, suppose  $p^2K/K^{2\alpha}=o(1)$ and Assumption \ref{ab} in Appendix A holds, then we have
	\begin{align}\label{theq2}
		\frac{\sqrt{K}v^{\top}(\hat{\theta}_{stage,1}-\theta^*)}{(v^{\top}\Sigma^{-1}\mathbb{E}[\mathbb{E}\{\dot{l}(X,Y,\theta^*)|\Delta(\theta^*)\}^{\otimes 2}]\Sigma^{-1}v)^{1/2}}\stackrel{d}{\to}\mathbf{N}(0,1),
	\end{align}
    where $\Sigma=\mathbb{E}\{\ddot{l}(X,Y,\theta^*)\}$.
\end{theorem}

\begin{remark}
Theorem \ref{thm2} establishes the necessary conditions governing the relationship between the parameter dimension and the number of data streams $K$  under the OSNI framework, while simultaneously determining the optimal checkpoint for initializing Newton's method. Notably, if the similarity between tasks increases as the sample size grows, the convergence rate of our estimator can be further improved.
In Theorem \ref{thm2}, we assume that $\theta_k$ on the $k$-th task is a constant-order random variable, in which case the convergence rate of our final estimator is of order $\sqrt{1/K}$. When $\theta_k$ converges to $\theta$ as $n$ increases, for example, if $\|\theta_k-\theta^*\|=O(n^{-\beta})$ with $0<\beta<1/2$ (assume the orders from $n_1$ to $n_k$ and $n$ are the same), then by simple Taylor's expansion, $\|\mathbb{E}\{\dot{l}(X,Y,\theta^*)|\Delta(\theta^*)\}^{\otimes 2}\|=O(n^{-2\beta})$, and from (\ref{theq2}) we know $\hat{\theta}_{stage,1}$ converges to $\theta^*$ with the rate $1/(\sqrt{K}n^{\beta})$.

\end{remark}

Theorem \ref{thm2} imposes overly stringent requirements on the proportion $\alpha$, necessitating the processor to simultaneously store complete information from a large number of data streams (exceeding $\sqrt{pK}$). For scenarios with substantial $K$ values, this approach fails to alleviate the computational and storage burdens on the processor. To address this limitation, we subsequently propose a multi-stage online Newton iteration method, which dramatically reduces the number of data streams required for computing the initial estimator while maintaining theoretical guarantees.


\subsection{Multi-stage Newton Iteration (MSNI)}

In this subsection, we propose a Multi-step Newton Iterative algorithm, in which the parameters are appropriately updated by Newton iterations. This process enables parameter calibration, preventing excessive deviation from the target values.

We can employ the OSNI as the initial phase of the Multi-Stage Newton Iteration (MSNI). In this first stage, Algorithm \ref{alg2} yields the preliminary estimate $\hat{\theta}_{stage,1}$ using $\lfloor K^{\alpha_1}\rfloor$ data streams, where $\lfloor K^{\alpha_1}\rfloor$ is intentionally kept small. For subsequent data streams, we dynamically update the gradient and Hessian matrix by evaluating them at the most recent parameter estimates, following the iterative update rule below:

\begin{equation}\label{3}
  \begin{aligned}	\hat{\theta}_{stage,t}=&\hat{\theta}_{stage,t-1}-\left[\frac{1}{\lfloor K^{\alpha_t}\rfloor}\left\{\sum_{k=1}^{\lfloor K^{\alpha_1}\rfloor}\ddot{L}_k(\hat{\theta}_{stage,0})\right.\right.\left.\left.+\sum_{s=2}^{t}\sum_{k=\lfloor K^{\alpha_{s-1}}\rfloor+1}^{\lfloor K^{\alpha_s}\rfloor}\ddot{L}_k(\hat{\theta}_{stage,s-1})\right\}\right]^{-1}
	\\&\times\frac{1}{\lfloor K^{\alpha_t}\rfloor}\left\{\sum_{k=1}^{\lfloor K^{\alpha_1}\rfloor}\dot{L}_k(\hat{\theta}_{stage,0})\right.\left.+\sum_{s=2}^{t}\sum_{k=\lfloor K^{\alpha_{s-1}}\rfloor+1}^{\lfloor K^{\alpha_s}\rfloor}\dot{L}_k(\hat{\theta}_{stage,s-1})\right\}, t=1,2,\cdots,T,
  \end{aligned}
\end{equation}

where $\alpha_i, i=1,2,\cdots,T$ represent fixed tuning parameters that we predefine within the range $(0,1)$. Under this iterative rule, the parameters are updated every $\lfloor K^{\alpha_{t}}\rfloor-\lfloor K^{\alpha_{t-1}}\rfloor$ batch of data, and as the parameters are updated iteratively and gradually stabilized, $\lfloor K^{\alpha_{t}}\rfloor-\lfloor K^{\alpha_{t-1}}\rfloor$ can become larger and larger, thus reducing the number of times the matrix inverse is computed. The specific algorithm is described in Algorithm \ref{alg3}.

\begin{algorithm}[ht]
   \caption{Multi-stage Newton Iteration(MSNI)}
   \label{alg3}
\begin{algorithmic}
   \STATE {\bfseries Input:} $\{X_{(k,i)},Y_{(k,i)} 
   \}_{i=1}^{n_k},k=1,\cdots,K$, $T$, $\alpha$, $\alpha_i$, $i=1,\cdots,T$.
   \STATE {\bfseries Output:} $\hat{\theta}_{stage,T}$.
   \STATE Obtain an estimate $\hat{\theta}_{stage,1}$ via Algorithm \ref{alg2}.
   \FOR{$t=2$ {\bfseries to} $T$}
      \FOR{$k=(\lfloor K^{\alpha_{t-1}}\rfloor+1)$ {\bfseries to} $\lfloor K^{\alpha_t}\rfloor$}
      \STATE Calculate and store $\dot{L}_k(\hat{\theta}_{stage,t-1})$. 
      \STATE Calculate and store $\ddot{L}_k(\hat{\theta}_{stage,t-1})$.
      \ENDFOR
   \STATE Calculate the one-step estimator $\hat{\theta}_{stage,t}$ by (\ref{3}).
\ENDFOR
\end{algorithmic}
\end{algorithm}
\begin{remark}

In the MSNI algorithm, the determination of each iteration's position is related to the randomness of the parameters. Due to the heterogeneity of the parameters and the inability to retain the data stream for an extended period, we cannot use the value of the Hessian matrix at the current latest parameter estimate in every iteration. Therefore, we aim to have a larger amount of data stream between each parameter update. This way, the influence of the low-precision parameter estimates obtained earlier on the estimation accuracy of the Hessian matrix can be negligible in the asymptotic sense, without affecting the limiting convergence rate and asymptotic normality of the estimators.
\end{remark}

 MSNI performs better when there is a comparatively large gap between the task parameters and requires less data for the initial estimation. Analogous to the convergence results in Theorem \ref{thm2}, Theorem \ref{thm3} provides corresponding theoretical guarantees for the estimator in Algorithm \ref{alg3}.

\begin{theorem}\label{thm3}
	Suppose that Assmuptions \ref{a11}-\ref{a6} hold. 
    Let $\alpha_0=\alpha$ in Algorithm \ref{alg3}. 
    If $p/K^{\alpha_0}=o(1)$ and $pK^{\alpha_t}/K^{2\alpha_{t-1}}=o(1)$  for $t=1,2,\cdots,T$,  we have
	$$\|\hat{\theta}_{stage,T}-\theta^*\|=O_p(\sqrt{p/K}).$$ Moreover, if $p^2K/K^{2\alpha_{T-1}}=o(1)$ and Assumption $\ref{ab}$  hold,
	then 
	\begin{align}\label{theq3}
		\frac{\sqrt{K}v^{\top}(\hat{\theta}_{stage,T}-\theta^*)}{\{v^{\top}\Sigma^{-1}\mathbb{E}(Z_k^{\otimes 2})\Sigma^{-1}v\}^{1/2}}\stackrel{d}{\to}\mathbf{N}(0,1),
	\end{align}
    where  $Z_k=\mathbb{E}\{\dot{l}(X_{(k,i)},Y_{(k,i)},\theta^*)|\Delta(\theta^*)=\theta_k-\theta^*\}$ and $\Sigma=\mathbb{E}\{\ddot{l}(X,Y,\theta^*)\}$.
\end{theorem}

\begin{remark}
Theorem \ref{thm3} extends our theoretical framework to incorporate Multi-step Newton-Raphson iterations, providing a complete characterization of both the estimator's convergence rate and the sufficient conditions guaranteeing asymptotic normality. The theorem's principal advancement lies in substantially relaxing the dimensional constraints (governing the relationship between parameter dimension $p$ and sample sizes $K^{\alpha_t}$) imposed by Theorem \ref{thm2}, thereby significantly expanding the algorithm's operational domain while maintaining theoretical guarantees. Moreover, similar to Theorem \ref{thm2}, if $\|\theta_k-\theta^*\|=O(n^{-\beta})$, we have $\mathbb{E}(Z_k^{\otimes 2})=O(n^{-2\beta})$ and then (\ref{theq3}) yields that $\hat{\theta}_{stage,T}$ converges to $\theta^*$ with the rate $1/(\sqrt{K}n^{\beta})$. This implies  when the similarity between tasks increases, the convergence rate of our estimator can become faster.
\end{remark}

Theorem \ref{thm3} establishes the consistency of  $\hat{\theta}_{stage,T}$ for $\theta^*$, which directly implies the ratio convergence, which directly implies the ratio convergence $F(\hat{\theta}_{stage,T})/F(\theta^*)\stackrel{P}{\to} 1$. This result demonstrates that the multi-stage estimator  $\hat{\theta}_{stage,T}$ asymptotically minimizes the global loss function $F(\cdot)$ to the same extent as the true parameter $\theta^*$.
Furthermore, the theorem reveals an asymptotic equi-impact property across data streams $\hat{\theta}_{stage,T}$. $\hat{\theta}_{stage,T}$ exhibits no systematic bias toward either earlier or later data streams in the iterative process. This property stems from the exponentially growing sample information incorporated at each iteration stage, which asymptotically dominates and cancels out the estimation bias from previous stages.

Our theoretical results naturally extend to the distributed calculation. The following theorem verifies that Algorithm \ref{alg3} maintains equivalent estimation efficiency - when i.i.d. samples are evenly distributed across streams, the estimator attains the same convergence rate as the distributed calculation.

\begin{theorem}\label{th4}
    Under Assmuptions \ref{a11}-\ref{a6}, suppose $\Delta (\theta) $ is a degenerate distribution, namely $\theta_1=\theta_2=\cdots=\theta^*$ and the distributions of $\{X_k,Y_k\}$ for $k=1,\cdots,K$ are the same. Define $\alpha_0=\alpha$ in Algorithm \ref{alg3}. 
     If $n_1=n_2=\cdots=n_K$, $p/(K^{\alpha_0}n)=o(1)$ and $pK^{\alpha_t}/K^{2\alpha_{t-1}}=o(1)$, 
then we have
	$$\|\hat{\theta}_{stage,T}-\theta^*\|=O_p(\sqrt{p/(Kn)}).$$ Moreover, if 
    Assumption $\ref{ac}$  holds,
	then 
	\begin{align*}
		\frac{\sqrt{Kn}v^{\top}(\hat{\theta}_{stage,T}-\theta^*)}{(v^{\top}\Sigma^{-1}\mathbb{E}\{\dot{l}(X,Y,\theta^*)^{\otimes 2}\}\Sigma^{-1}v)^{1/2}}\stackrel{d}{\to}\mathbf{N}(0,1),
	\end{align*}
    where $\Sigma=\mathbb{E}\{\ddot{l}(X,Y,\theta^*)\}$.
\end{theorem}

\begin{remark}
Theorem \ref{th4} establishes that when i.i.d. samples are uniformly distributed across heterogeneous data streams, the aggregated statistic computed by Algorithm \ref{alg3} attains asymptotic convergence rates matching those of distributed M-estimation frameworks (e.g., \cite{jordan2019communication}). This stands in contrast to the rate limitation ($p/K$) characterizing the estimator in Theorem \ref{thm2}, a fundamental consequence of cross-stream parameter heterogeneity. These results collectively demonstrate that our methodology preserves theoretical optimality in convergence rate under uniform sampling conditions. Specifically, 
we can also interpret Theorem \ref{thm3} as the scenario where the similarity between tasks increases sufficiently rapidly with the growing sample size,  that is, $\|\theta_k-\theta^*\|=o(1/\sqrt{n})$. 
\end{remark}


\section{Experiments}

\subsection{Simulation Studies}

In this subsection, we perform comprehensive simulation studies to assess the empirical performance of our proposed MSNI method. We examine two fundamental statistical models: linear regression model and logistic regression model. For each model specification, we conduct 100 repeated experiments to ensure reliable performance evaluation. All reported results represent the averaged outcomes across these replications. For comparative analysis, we conducted parallel experiments using several benchmark methods, including: Weighted Least Squares Estimation (WLSE) and two classical continual learning methods.

According to the framework of Regularization-based Continual Learning(RBCL) approaches in \citet{wu2024meta}, we adopt the following iterative approach for estimation:
$$\theta_k:=\theta_{k-1}-\alpha\left(\sum_{i=1}^{j-1} \ddot{L}^i\right)^{-1} \dot{L}^j(\theta),$$
where $\ddot{L}^i$ represents the Fisher information matrix obtained from training the $i$-th task, $\dot{L}^j(\theta)$ represents the gradient on the $j$-th task. Furthermore, we employ the Gradient Episodic Memory (GEM) \citep{lopez2017gradient} algorithm, which leverages both memory and gradient information, as a comparative baseline in our study.

We define $p$ as the dimension of parameter vector ${\theta}$, with $K$ representing the total number of batches and $n_k, k=1,2,\cdots, K$ denoting the size of each data batch. We set the parameter value as ${\theta}_0=10\times\frac{p^{-1/2}(p-i)}{p-1}$ and consider the following two parameter settings:

\textbf{Setting 1:} The true parameter for each batch of data is
$$\theta_k=\theta_0+\eta_k,k=2,\cdots,K,$$
where $\eta_k\sim N(0,\sigma^2)$. For the estimations $\hat{\theta}_k$ after obtaining the $k$-th batch of data, we calculate
$MSE_k=\Vert \hat{\theta}_k-\theta_0\Vert^2 $
to evaluate the performance of the algorithms. 

\textbf{Setting 2:} Different tasks have different true parameters as
$$\theta^j=\theta_0+\eta^j, j=1,\cdots,M=5,$$
where $M$ is the total number of tasks. Each task's data uses the online setting and is inputted in batches. Let 
$$MSE_{k}^{j}=\Vert \hat{\theta}_k-\theta^j\Vert^2 \ {\rm{and}}\ MMSE_k^i=\frac{1}{i}\sum\limits_{j=1}^iMSE_{k}^{i,j}, j\le i, $$
where $i$ denotes the current task index. Thus, $MSE_k^j$ quantifies the discrepancy between the parameters estimated after the $k$-th data batch and the truth parameter of the $j$-th task, while $MMSE_k^i$ represents the average of such $MSE_k^j$ values across all tasks learned so far. 

We also denote $\Vert \theta_k-\theta_0\Vert^2$ or $\Vert \theta^j-\theta_0\Vert^2$ as deviation. It can be used to measure the size of changes in real parameters. In Setting 1, we treat the parameters as fully random variables where each data batch possesses distinct true parameters. This configuration provides a more rigorous validation of our theoretical results. Setting 2 aligns with conventional continual learning paradigms, maintaining fixed true parameters for each task. We conduct experiments under both configurations to comprehensively verify both our theoretical 
results and algorithmic efficacy. 

\begin{figure}[htb!]
		\centering
		\includegraphics[width=7cm,height=5cm]{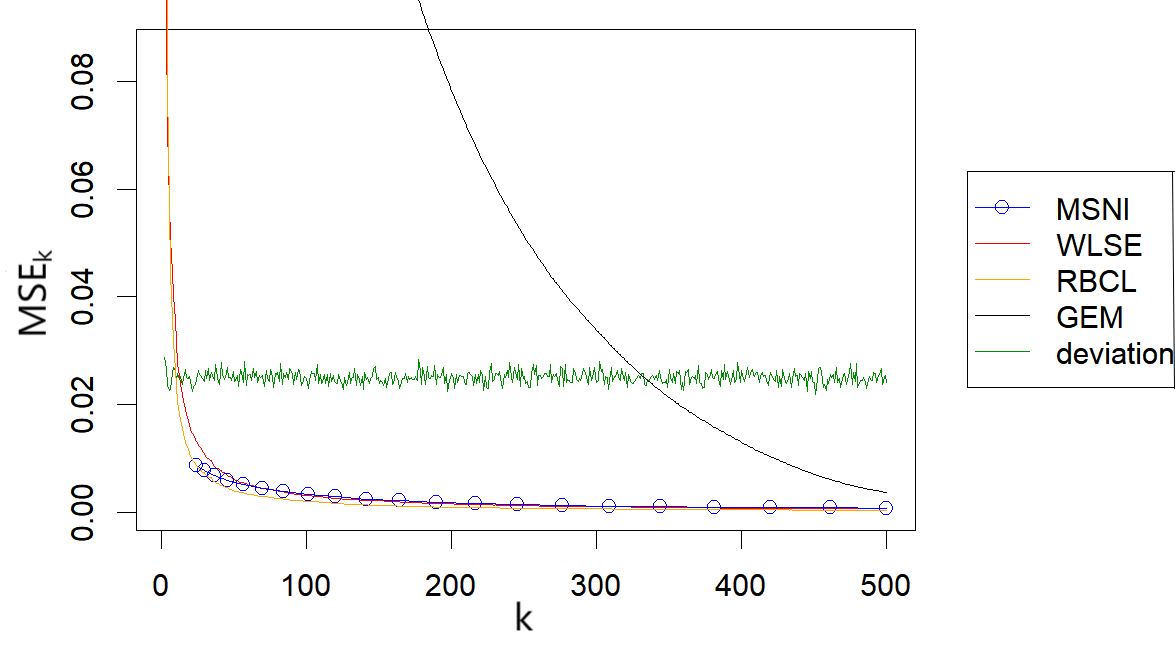}
		\quad
		\centering
		\includegraphics[width=7cm,height=5cm]{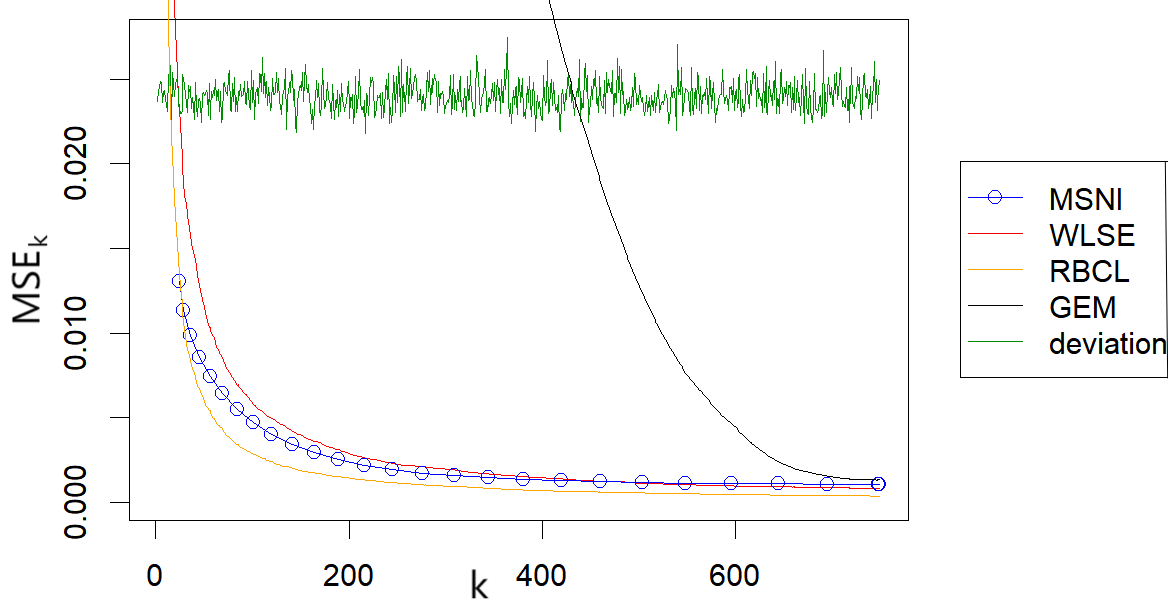}
		
		\centering
		\includegraphics[width=7cm,height=5cm]{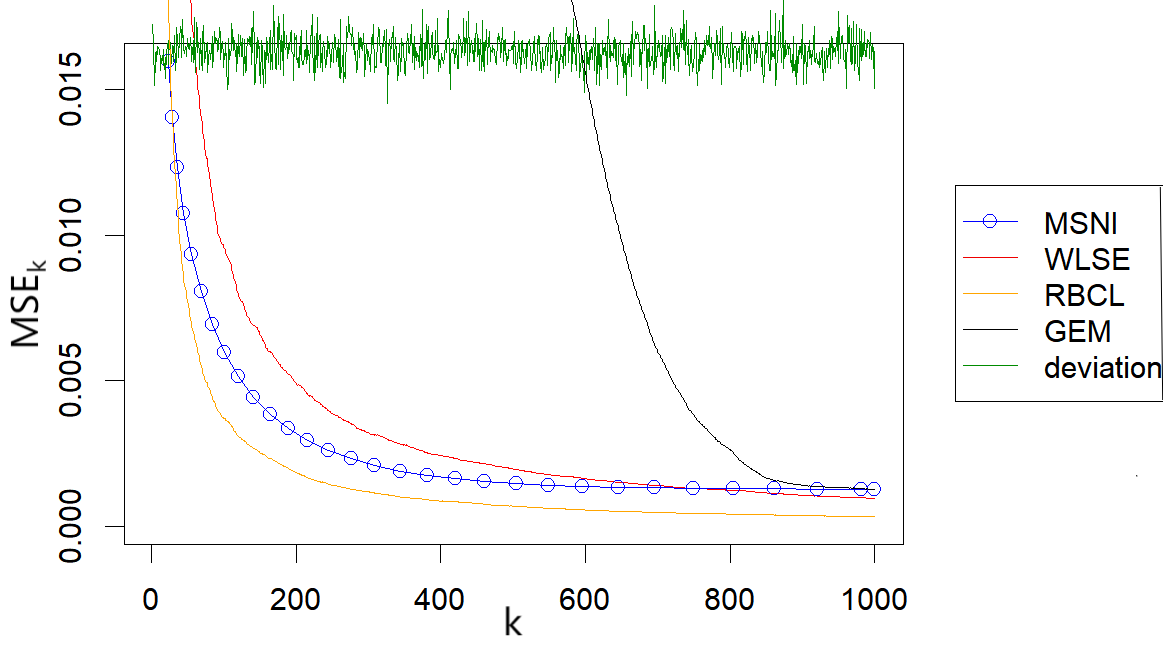}
		\quad
		\centering
		\includegraphics[width=7cm,height=5cm]{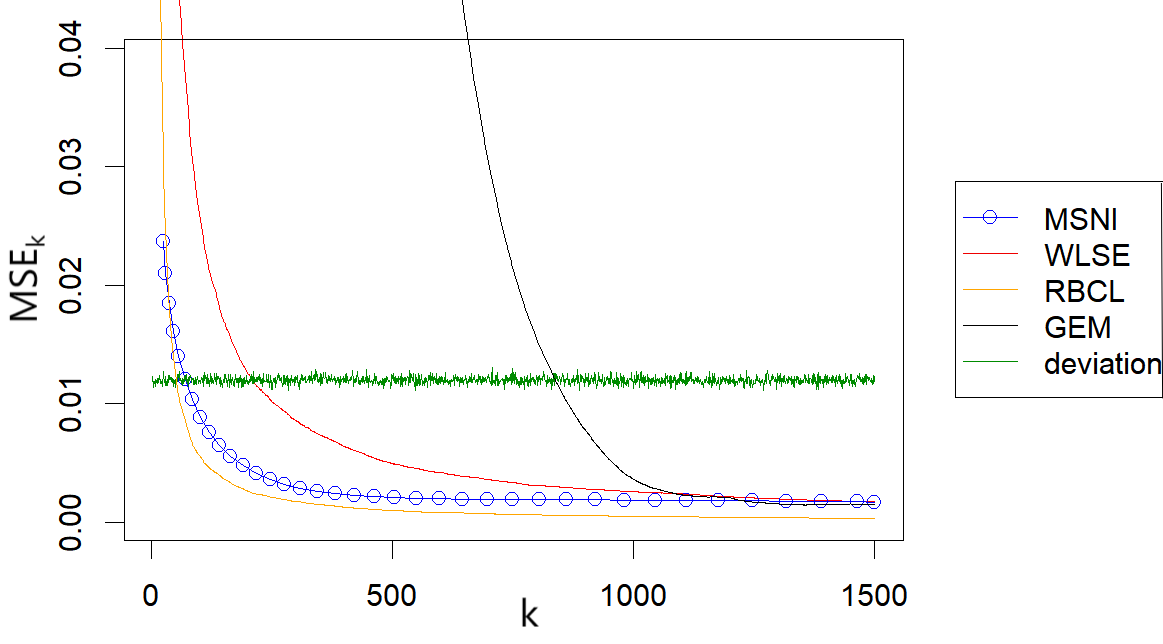}
		\vspace{-0.25cm}
		\caption{The experimental results of the linear regression model under Setting 1, which is set as follows (a) $K=500, n_k=100, p=10$, $\sigma=1/20$; (b) $K=750, n_k=100, p=15$, $\sigma=1/25$; (c) $K=1000, n_k=100, p=20$, $\sigma=1/35$; (d) $K=1500, n_k=100, p=30$, $\sigma=1/50$.}\label{linear1}
	\end{figure}

\begin{figure}[htb!]
		\centering
		\includegraphics[width=7cm,height=5cm]{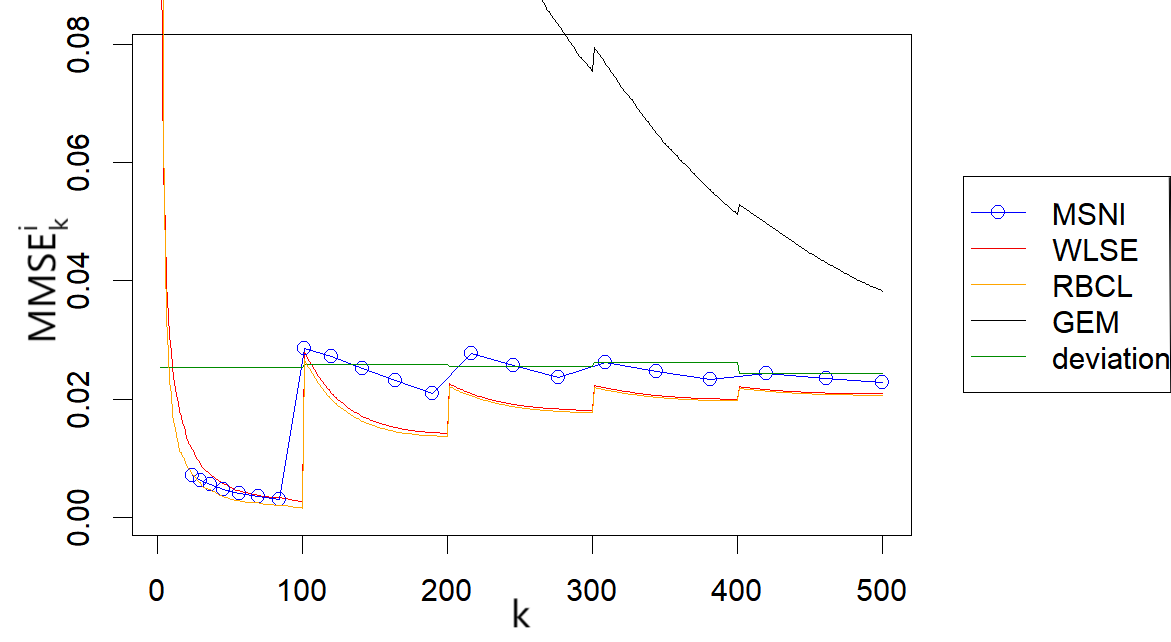}
		\quad
		\centering
		\includegraphics[width=7cm,height=5cm]{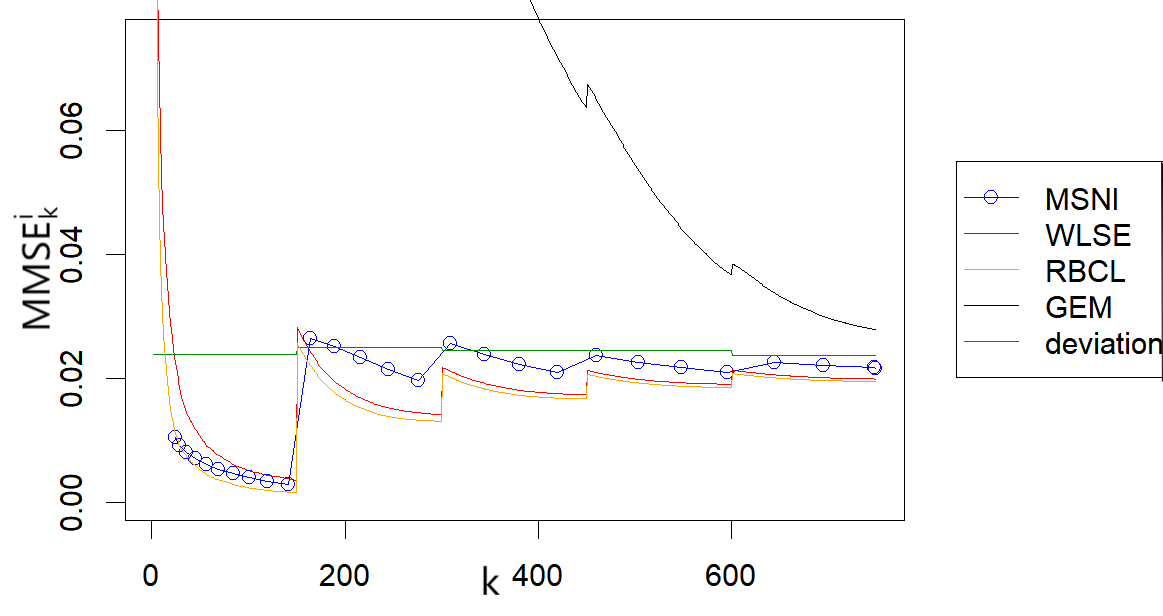}
		
		\centering
		\includegraphics[width=7cm,height=5cm]{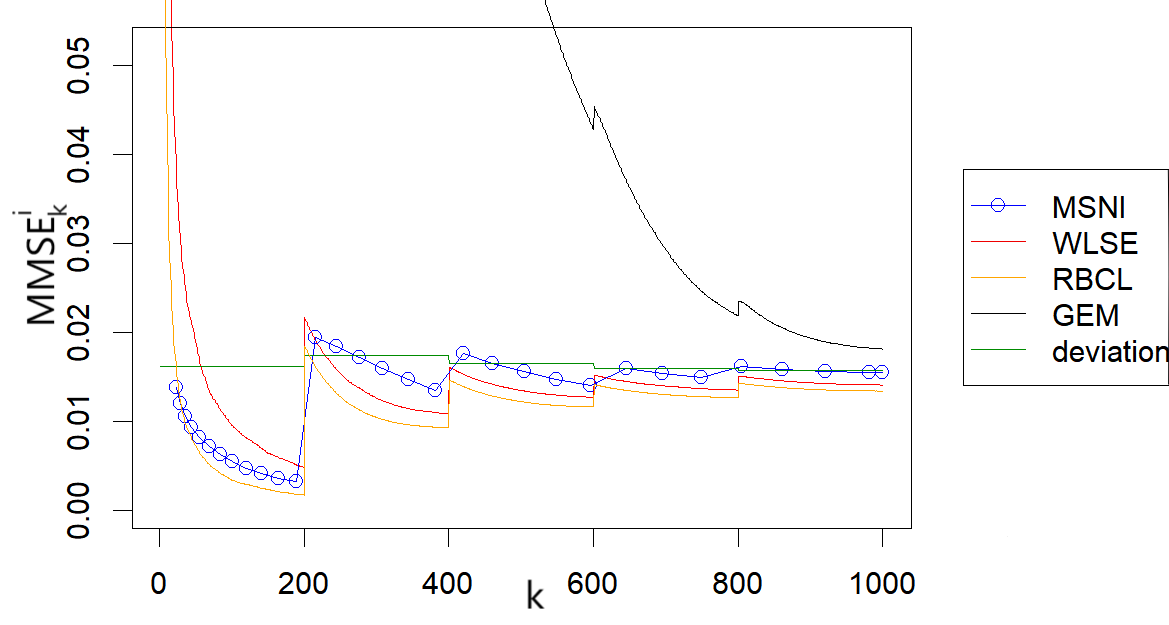}
		\quad
		\centering
		\includegraphics[width=7cm,height=5cm]{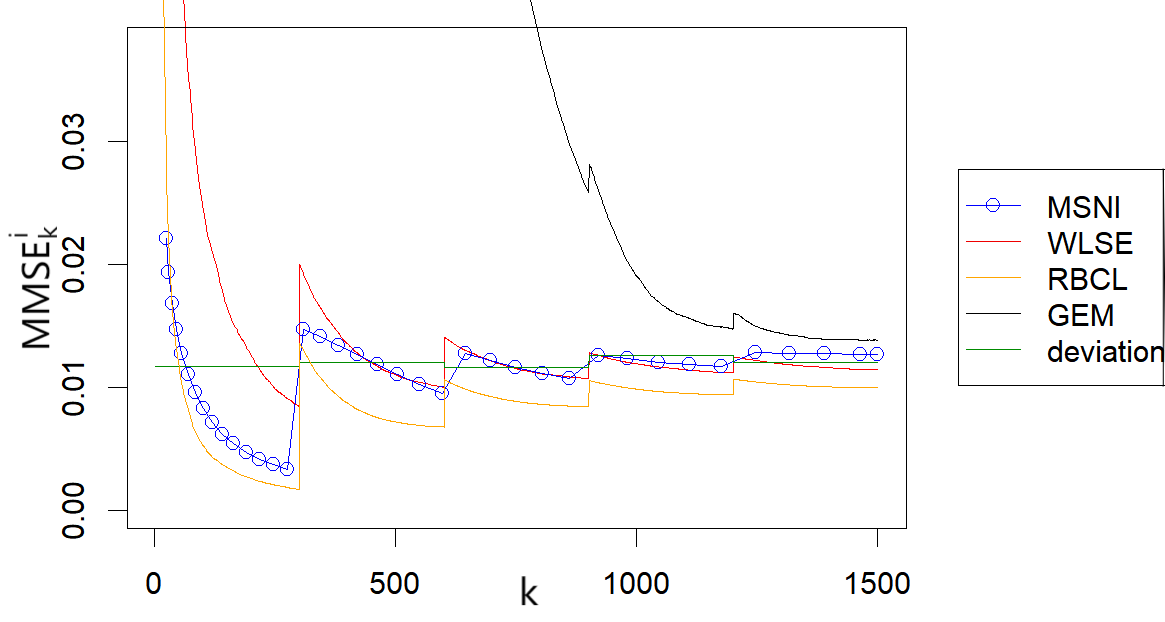}
		\vspace{-0.25cm}
		\caption{The experimental results of the linear regression model under Setting 1, which is set as follows (a) $K=500, n_k=100, p=10$, $\sigma=1/20$; (b) $K=750, n_k=100, p=15$, $\sigma=1/25$; (c) $K=1000, n_k=100, p=20$, $\sigma=1/35$; (d) $K=1500, n_k=100, p=30$, $\sigma=1/50$.}\label{linear2}
	\end{figure}

$\mathbf{Case 1: Linear }$  $\mathbf{Regression.}$	
We generate datasets from the linear regression model as: $y=X\theta+\varepsilon$, where $X\sim N(\boldsymbol{0},\Sigma)$ with $\Sigma=(\Sigma_{ij})_{p\times p}$ and $\Sigma_{ij}=0.5^{|i-j|}$ for $1\le i,j \le p$. Figure \ref{linear1} and Figure \ref{linear2} respectively present  the parameter estimation of the proposed algorithm for this model under two different settings.

\begin{figure}[htb!]
		\centering
		\includegraphics[width=7cm,height=5cm]{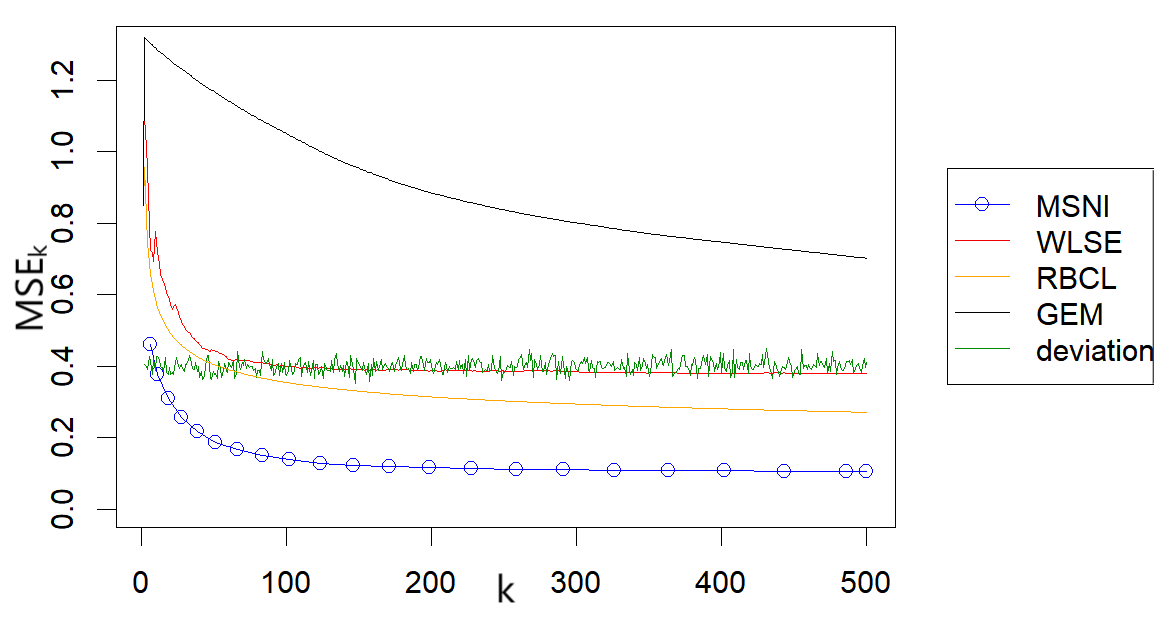}
		\quad
		\centering
		\includegraphics[width=7cm,height=5cm]{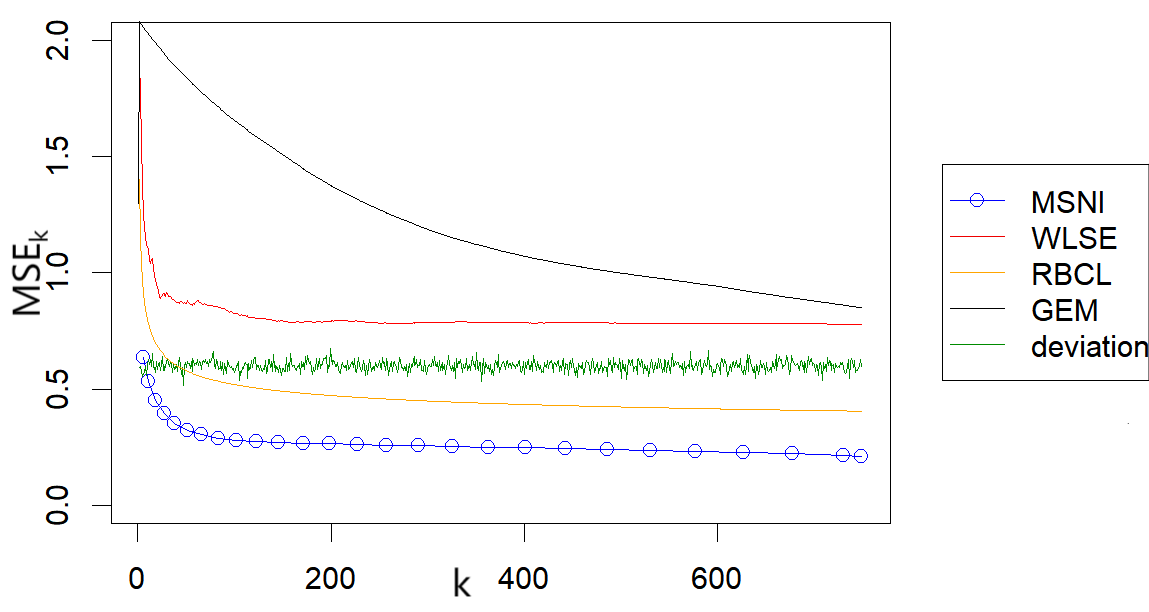}
		
		\centering
		\includegraphics[width=7cm,height=5cm]{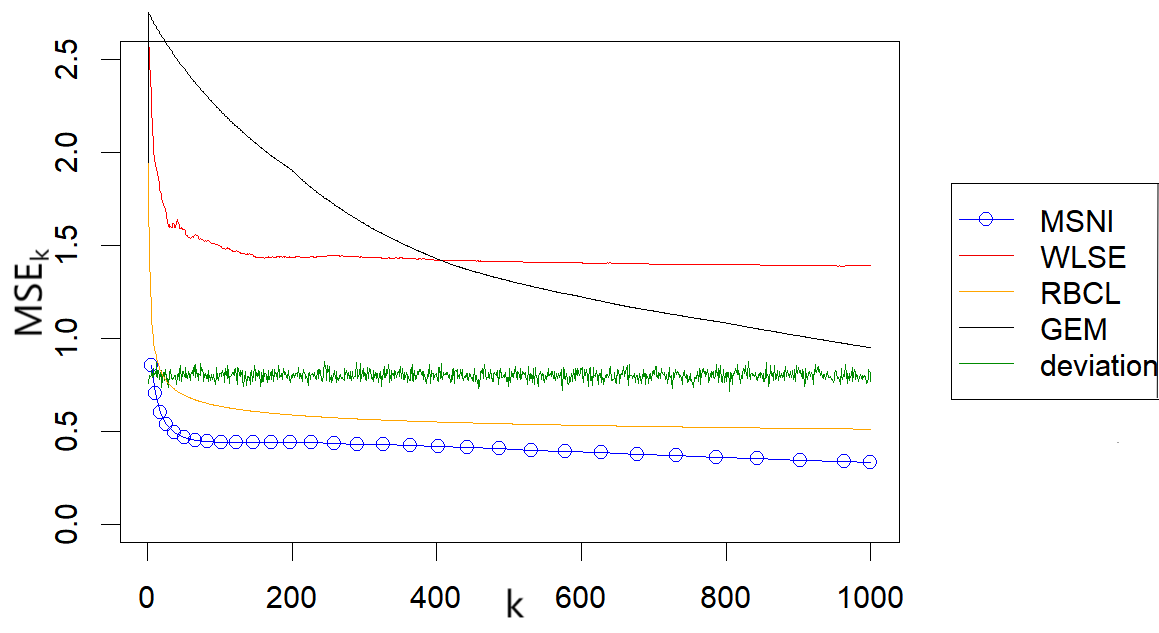}
		\quad
		\centering
		\includegraphics[width=7cm,height=5cm]{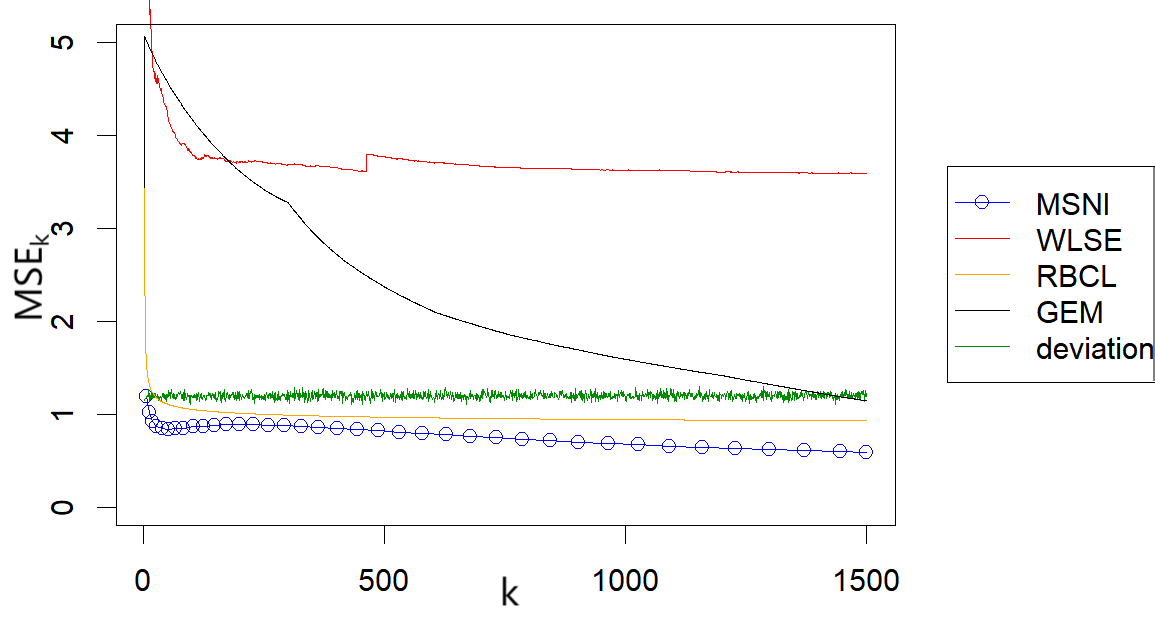}
		\vspace{-0.25cm}
		\caption{The experimental results of the logistic regression model under Setting 1, which is set as follows (a) $K=500, n_k=1000, p=10$, $\sigma=1/5$; (b) $K=750, n_k=1000, p=15$, $\sigma=1/5$; (c) $K=1000, n_k=1000, p=35$, $\sigma=1/5$; (d) $K=1500, n_k=1000, p=30, \sigma=1/5$.}\label{logist1}
	\end{figure}

\begin{figure}[htb!]
		\centering
		\includegraphics[width=7cm,height=5cm]{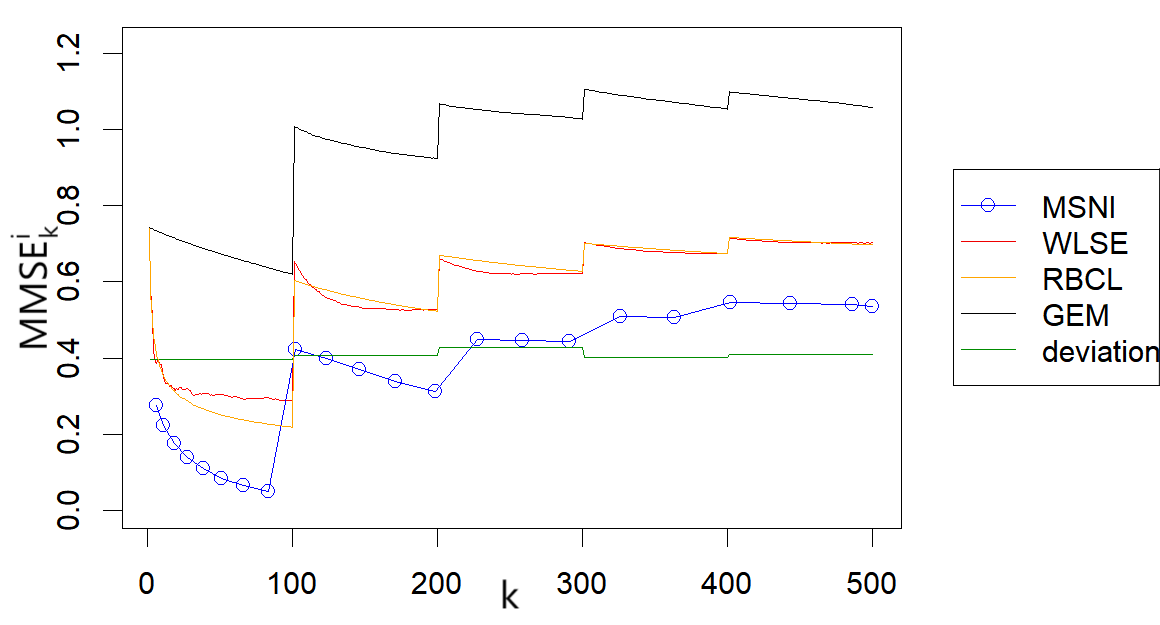}
		\quad
		\centering
		\includegraphics[width=7cm,height=5cm]{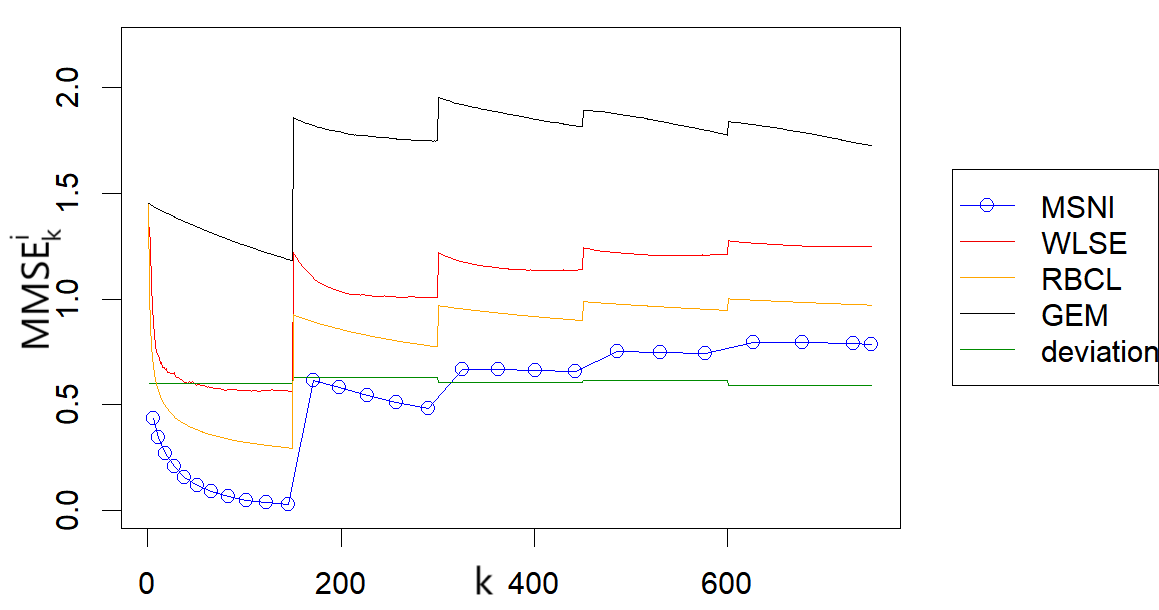}
		
		\centering
		\includegraphics[width=7cm,height=5cm]{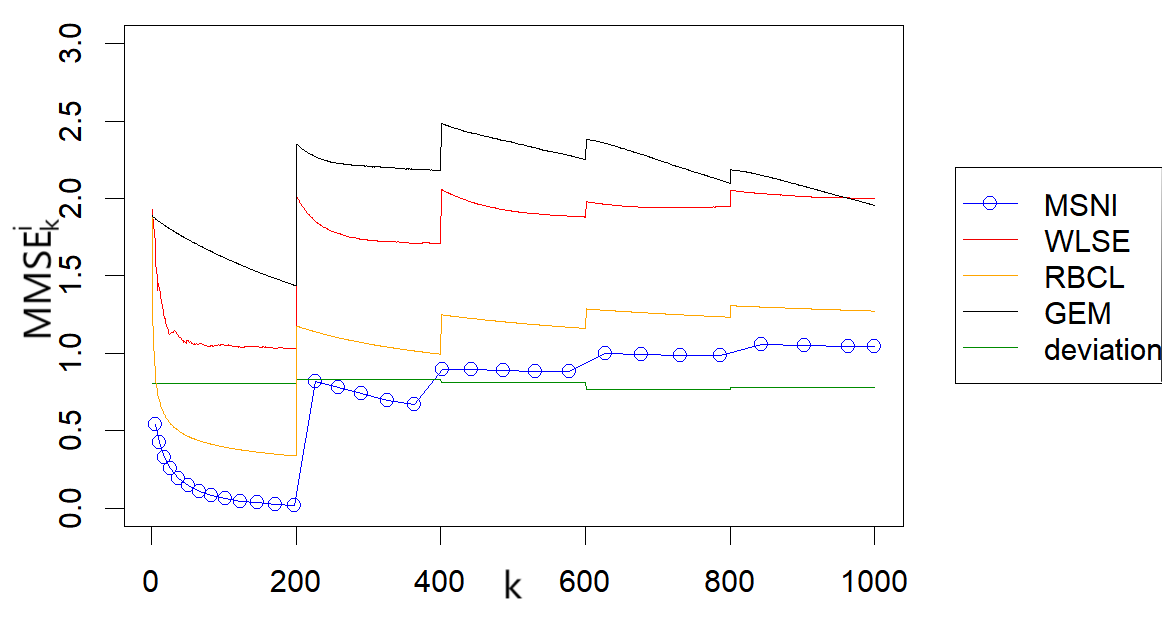}
		\quad
		\centering
		\includegraphics[width=7cm,height=5cm]{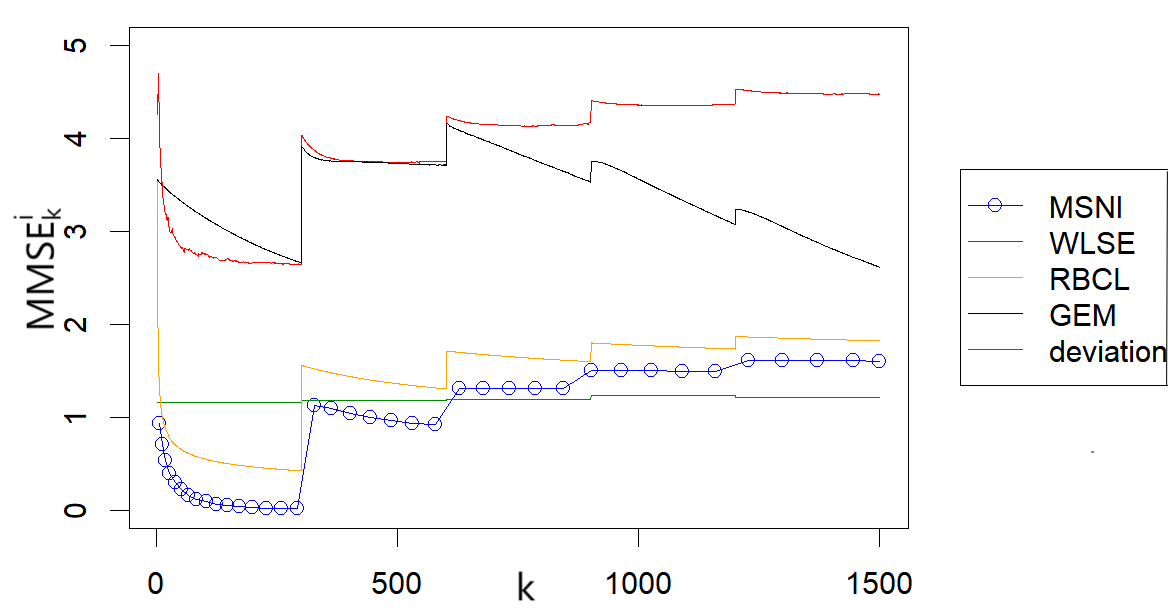}
		\vspace{-0.25cm}
		\caption{The experimental results of the logistic regression model under Setting 1, which is set as follows (a) $K=500, n_k=1000, p=10$, $\sigma=1/5$; (b) $K=750, n_k=1000, p=15$, $\sigma=1/5$; (c) $K=1000, n_k=1000, p=35$, $\sigma=1/5$; (d) $K=1500, n_k=1000, p=30, \sigma=1/5$.}\label{logist2}
	\end{figure}

$\mathbf{Case 2: Logistic }$  $\mathbf{Regression.}$ Generate data from the following logistic regression model: $$P(y=1|X,\theta)=\frac{\exp(X^T\theta)}{1+\exp(X^T\theta)}.$$ 
Here we set the distribution of $X$ to be the same as in Case 1.
Figure \ref{logist1} and Figure \ref{logist2} respectively show the performance of the algorithm under two different settings of the logistic regression model.

For linear regression models, the Hessian matrix computation depends exclusively on the data itself and is independent of parameters. With sufficient iterations, all algorithms converge to similar results. In logistic regression models, the MSNI algorithm demonstrates superior performance, significantly outperforming the WLSE and GEM algorithms, which show relatively poor results. Notably, as dimensionality increases, more data batches are required to achieve optimal performance, which aligns perfectly with our theoretical results. Overall, the MSNI algorithm exhibits the best comprehensive performance across all experiments.

\subsection{Real Data}

In this subsection, we analyze the MNIST dataset of handwritten digits and CIFAR-10. The MNIST dataset is a large database of handwritten numbers, widely used for machine learning related training and testing. It has a training set of 60,000 examples, and a test set of 10,000 examples. The covariate of the MNIST dataset is a 28 by 28 pixel image, and the response variable is s from 0 to 9. 
CIFAR-10 is a classic image classification dataset containing 60,000 RGB color images of 10 categories, with 6,000 images per category. This dataset 
is widely used to test deep learning algorithms and evaluate benchmark performance in the field of computer vision.

We adopt the settings of DIL. The ten categories of images are divided into two main categories. We set a total of five tasks, each of which is a binary classification task with non-overlapping data included in each task. We extract features from the two datasets using the VGG16 neural network, resulting in 512-dimensional data that can be used for logistic regression model. We also use online data settings, where data is divided into small batches and sequentially inputted for training. 

In this section, we use classical evaluation metrics including Average Incremental Accuracy(AIA) \citep{hou2019learning, douillard2020podnet}, Forward Transfer(FWT) and Backward Transfer(BWT) \citep{lopez2017gradient} to test the performance of the algorithm. Let $R_{i,j}$ denote the testing accuracy of the model on task $j,1\le j \le i$, after being trained on task $i,1\le i \le M$. $R_{0,j}$ is the testing accuracy of a model randomly initialized on task $i$. The Average Accuracy \citep{chaudhry2018riemannian} of testing after training on task $i$ is
$AA_i=\frac{1}{i}\sum\limits_{j=1}^iR_{i,j}, $
and the Average Incremental Accuracy is defined as $AIA=\frac{1}{M}\sum\limits_{i=1}^TAA_i.$ The Forward Transfer and Backward Transfer are defined as: $FWT=\frac{1}{T-1}\sum\limits_{i=2}^T R_{i-1,i}-R_{0,i}$ and $BWT= \frac{1}{T-1}\sum\limits_{i=1}^{T-1}R_{T,i}-R_{i,i}$. FWT is used to measure the impact of existing knowledge on the performance of subsequent tasks. BWT reflects the forgetting of previous tasks caused by subsequent tasks. 

\begin{figure}[htb!]
		\centering
		\includegraphics[width=8cm,height=6cm]{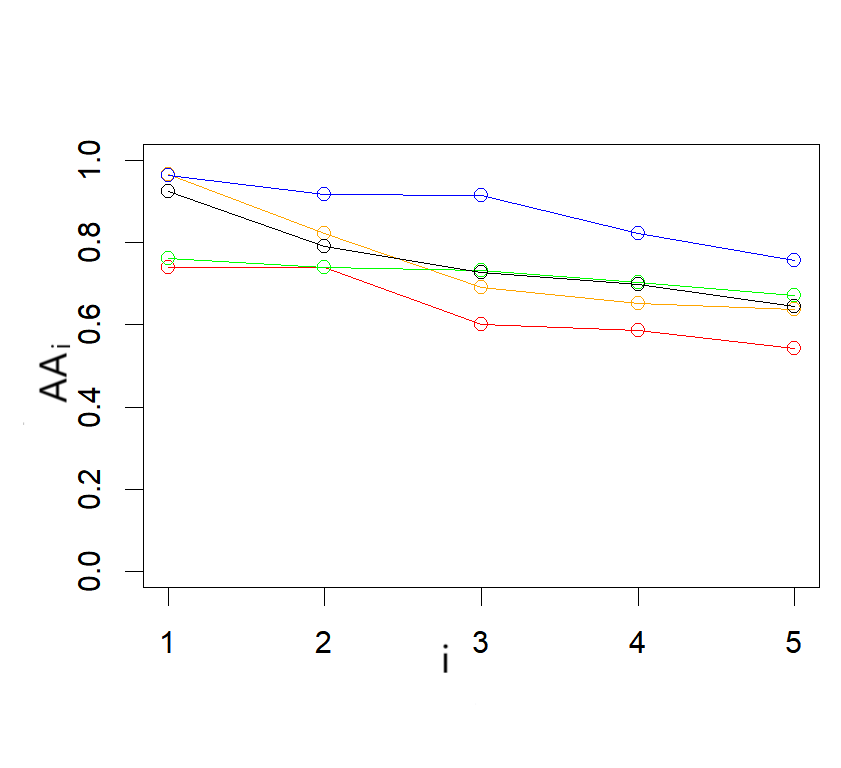}
		\quad
		\centering
		\includegraphics[width=8cm,height=6cm]{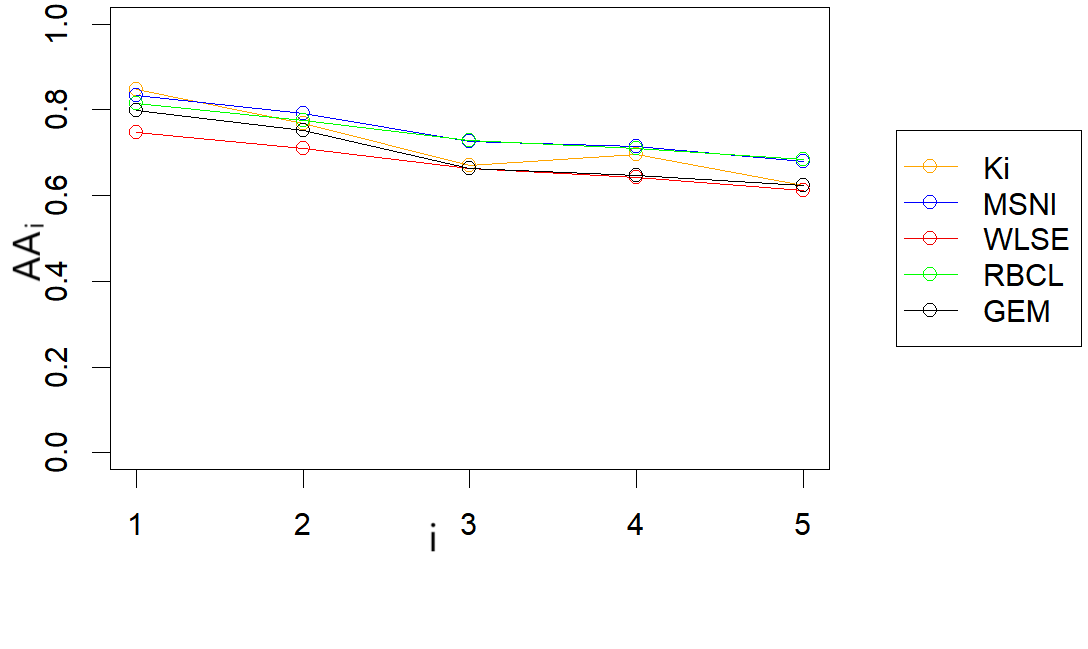}
		\vspace{-0.25cm}
		\caption{(a)The Average Accuracy tested on the MNIST dataset; (b)The Average Accuracy tested on the CIFAR-10 dataset}\label{real}
	\end{figure}

Figure \ref{real} illustrates the Average Accuracy of the algorithms after training on each task for the MNIST and CIFAR-10 datasets as new tasks are continuously added. Table \ref{TMNIST} and Table \ref{TCIFAR} respectively show the overall performance, the learning plasticity of new knowledge and the memory stability of old knowledge tested on the two datasets.

\begin{table}[t]
\caption{MNIST}
\label{TMNIST}
\vskip 0.1in
\begin{center}
\begin{small}
\begin{sc}
\begin{tabular}{lcccccr}
\toprule
 & MLE & MSNI & WLSE &RBCL &GEM \\
\midrule
FWT&-0.3049&-0.4067&-0.4046&-0.4204&-0.4050\\
BWT&-0.9836&-0.0387&-0.1337&-0.0444&0.0076\\
AIA&0.7531&0.8745&0.6421&0.7219&0.7567\\
\bottomrule
\end{tabular}
\end{sc}
\end{small}
\end{center}
\vskip -0.1in
\end{table}

\begin{table}[t]
\caption{CIFAR-10}
\label{TCIFAR}
\vskip 0.1in
\begin{center}
\begin{small}
\begin{sc}
\begin{tabular}{lcccccr}
\toprule
 & MLE & MSNI & WLSE &RBCL &GEM \\
\midrule
FWT&-0.0729&-0.1349&-0.1514&-0.1155&-0.1611\\
BWT&-0.7844&0.0090&-0.0071&-0.0425&0.0017\\
AIA&0.7211&0.7492&0.6749&0.7421&0.6969\\
\bottomrule
\end{tabular}
\end{sc}
\end{small}
\end{center}
\vskip -0.1in
\end{table}

In our implementation, we set the step factors to 1.0, 0.1, and 0.01 for each algorithm respectively, and report the optimal results obtained from these configurations. Across classification tasks, the MNIST dataset exhibits significant variation, where the MSNI algorithm demonstrates the most pronounced improvement. In contrast, for the more homogeneous classification tasks in CIFAR-10, the performance differences between algorithms diminish, yielding closely comparable results. Nevertheless, our proposed algorithm consistently achieves the best overall performance. 

\section{Conclusion}\label{sec-conc}

In this paper, we present a statistical analysis of the continual learning problems and propose a statistical framework for addressing the general continual learning challenge, with the aim of mitigating the issue of catastrophic forgetting. We propose the Multi-Step Newton Iteration (MSNI) algorithm, which is capable of processing online data efficiently. Our algorithm can adapt to specific requirements by reducing both the number of iterations and the frequency of calculating the inverse of the Hessian matrix, thus alleviating computational pressure. We establish the weakly normal limit distribution for the estimator,  which can be further used for statistical inference. As the volume of data increases, we allow the dimension of parameters to diverge to infinity. In the absence of heterogeneity in the data stream, our algorithm attains a convergence rate comparable to that of distributed computing methods. We apply the algorithm to the MNIST and CIFAR-10 datasets to verify its effectiveness.

Several promising research directions remain open for future exploration. First, our method requires the computation of the Hessian matrix, which necessitates the loss function being twice differentiable. This poses challenges when dealing with quantile regression or loss functions that violate this condition (e.g., absolute error loss). To address this limitation, we suggest incorporating smoothing techniques such as kernel smoothing functions \citep{horowitz1992smoothed}. Second, to further reduce computational overhead, we also consider incorporating alternative approximation techniques for Hessian matrices \citep{wang2024distributed}. Moreover, the proposed method might potentially be used to handle high-dimensional data by incorporating penalty functions, when sparsity exists in the data structure. Additionally, we propose to explore the potential of extending our current model to a semiparametric or nonparametric framework, drawing inspiration from existing online learning methodologies in these domains \citep{fang2023online, yang2024online}, thereby developing more versatile modeling approaches with broader applicability. Finally, we also consider employing relevant methods within the Reproducing Kernel Hilbert Space (RKHS) to process online data streams \citep{bouboulis2017online}, while providing a meticulous theoretical analysis.









\bibliographystyle{elsarticle-harv}
\bibliography{bibliography.bib}

\end{document}